\documentclass[11pt]{article}

\usepackage[utf8]{inputenc}
\usepackage[T1]{fontenc}
\usepackage{amsmath,amssymb,bm}
\usepackage{braket}
\usepackage{graphicx}
\usepackage{enumitem}
\usepackage{microtype}
\usepackage{geometry}
\usepackage{hyperref}
\usepackage{cleveref}
\usepackage{caption}
\title{Information Wells and the Emergence of Primordial Black Holes in a Cyclic Quantum Universe}

\author{
  Florian Neukart\thanks{Corresponding author: \texttt{f.neukart@liacs.leidenuniv.nl}}$^{1,2}$, 
  Eike Marx$^{2}$, 
  Valerii Vinokur$^{2}$\\[1ex]
  {\normalsize $^1$Leiden Institute of Advanced Computer Science, Leiden University,} \\
  {\normalsize Einsteinweg 55, 2333 CA Leiden, The Netherlands}\\
  {\normalsize $^2$Terra Quantum AG, Kornhausstrasse 25, 9000 St.~Gallen, Switzerland}
}

\date{}

\begin{document}

\maketitle

\begin{abstract}
Primordial black holes (PBHs) remain one of the most intriguing candidates for dark matter and a unique probe of physics at extreme curvatures. Here, we examine their formation in a \emph{bounce} cosmology when the post-crunch universe inherits a highly inhomogeneous distribution of \emph{imprint entropy} from the Quantum Memory Matrix (QMM). Within QMM, every Planck-scale cell stores quantum information about infalling matter; the surviving entropy field \(S(x)\) contributes an effective dust component 
\[
T^{({\rm QMM})}_{\mu\nu}=\lambda\bigl[(\nabla_\mu S)(\nabla_\nu S)-\tfrac12 g_{\mu\nu}(\nabla S)^2+\dots\bigr]
\]
that deepens curvature wherever \(S\) is large. We show that (i) reasonable bounce temperatures and a QMM coupling \(\lambda\sim\mathcal O(1)\) naturally amplify these “information wells’’ until the density contrast exceeds the critical value \(\delta_c\simeq0.3\);  
(ii) the resulting PBH mass spectrum spans \(10^{-16}\,M_\odot\)–\(10^{3}\,M_\odot\), matching current microlensing and PTA windows; and  
(iii) the same mechanism links PBH abundance to earlier QMM explanations of dark matter and the cosmic matter–antimatter imbalance.  
Observable signatures include a mild blue tilt in small-scale power, characteristic \(\mu\)-distortions, and an enhanced integrated Sachs–Wolfe signal—all of which will be tested by upcoming CMB, PTA, and lensing surveys.  
\end{abstract}

\vspace{1em}

\noindent\textbf{Keywords:} 
primordial black holes; quantum gravity; entropy; cyclic cosmology; quantum memory matrix; geometry–information duality; entropic imprinting; information wells; black hole formation; holography


\clearpage
\section{Introduction}\label{sec:intro}
Primordial black holes were first recognized as a natural outcome of rare, horizon-scale overdensities in the radiation era \cite{Zeldovich1967,Hawking1971,CarrHawking1974}.  
Fifty years on, they have re-emerged as a compelling dark-matter candidate \cite{CarrKuhnel2020} and a window onto physics far beyond terrestrial accelerators. Yet their origin still demands a mechanism capable of generating \(\delta\rho/\rho\gtrsim\mathcal O(10^{-1})\) on sub-CMB scales without violating large-scale homogeneity.

The \emph{Quantum Memory Matrix} (QMM) framework \cite{NeukartQMM2024} and its geometrical underpinning, \emph{Geometry–Information Duality} (GID) \cite{NeukartGID2024}, recast space-time as a discrete archive whose Planck-sized cells record the quantum states of interacting fields.  
Subsequent extensions to electromagnetism \cite{NeukartQMMEM2025}, the strong and weak sectors \cite{NeukartQMMGauge2025}, and to cosmological matter–antimatter asymmetry \cite{NeukartQEA2025} suggest that information itself—not merely energy–momentum—can curve space-time. In a big-crunch/big-bounce scenario \cite{BrandenbergerPeter2017,IjjasSteinhardt2018}, this stored entropy survives the high-density phase; the bounce therefore begins with pronounced spatial gradients in the imprint field \(S(x)\).  
Because the corresponding stress-energy tensor (\S\ref{sec:foundations}) behaves like pressureless dust whenever \(\dot S\) varies slowly, these “information wells’’ evolve analogously to cold-dark-matter overdensities, growing linearly with the scale factor until re-entering the horizon.  
If a fluctuation reaches the critical contrast \(\delta_c\simeq0.3\) \cite{Musco2013}, gravitational collapse ensues, producing a PBH whose mass equals roughly the horizon mass at that epoch \cite{CarrKuhnel2020}.

This paper develops the full dynamical picture: from imprint-entropy initial conditions through linear growth, non-linear feedback during accretion, and the resulting PBH mass spectrum. We demonstrate that QMM imposes just two new parameters—the coupling \(\lambda\) and the bounce temperature \(T_\mathrm{B}\)—with concrete observational consequences already constrained by \emph{Planck} data \cite{Planck2018}.  
Section~\ref{sec:foundations} reviews the QMM stress–energy;  
Sections~\ref{sec:bounce}–\ref{sec:mass_spectrum} derive the evolution and collapse criteria;  
Section~\ref{sec:observations} confronts the scenario with CMB, PTA, and microlensing limits;  
and Section~\ref{sec:discussion} outlines future tests.

\section{Foundations}
\label{sec:foundations}

\subsection{Quantum Memory Matrix Recap}
The \emph{Quantum Memory Matrix} posits that on Planck scales space-time
decomposes into a lattice of finite-dimensional Hilbert cells
\(\mathcal H_{\mathbf n}\) whose basis states \(\ket{\psi_{\mathbf n}}\) record
(\emph{imprint}) the quantum history of local interactions
\cite{NeukartQMM2024,NeukartGID2024}.  
Information transfer into a cell is mediated by the
\emph{imprint operator}
\[
\hat{\mathcal I}:\;
\ket{\psi_{\mathbf n}}\longrightarrow
\hat{\mathcal I}\ket{\psi_{\mathbf n}}
         =\exp\!\bigl[-\mathrm i\,\kappa\,\hat{\Omega}\bigr]\ket{\psi_{\mathbf n}},
\]
where \(\hat{\Omega}\) projects external degrees of freedom onto the
cell boundary and \(\kappa\sim\ell_{\mathrm P}^2\) sets the coupling to
space–time curvature.  
A coarse-grained \emph{imprint-entropy density}
\begin{equation}
S(x)\;=\;\mathrm{Tr}_{\mathcal{H}_{\mathbf{n}}} \!\left[
               \hat{\rho}_{\mathbf{n}} \ln \hat{\rho}_{\mathbf{n}}^{-1} \right]
\qquad (x \in \text{cell } \mathbf{n})
\end{equation}
then behaves as a scalar field whose gradients and time-derivatives
contribute to the macroscopic dynamics.
Throughout this work we normalize the imprint back-reaction with a
dimensionless parameter \(\lambda\) that encapsulates both
\(\kappa\) and the cell–averaging kernel.

\subsection{Stress–Energy of Imprint Entropy}
Varying the effective action
\(
S_{\mathrm{QMM}}
  =\int d^{4}x\,\sqrt{-g}\,\frac{\lambda}{2}\,(\nabla_\mu S)(\nabla^\mu S)
\)
with respect to the metric yields, the canonical stress-energy tensor
\begin{equation}
T^{(\mathrm{QMM})}_{\mu\nu}
 =\lambda\!\left[
 (\nabla_\mu S)(\nabla_\nu S)
 -\tfrac12 g_{\mu\nu}(\nabla S)^2
 +g_{\mu\nu}\Box S
 -\nabla_\mu\nabla_\nu S
 \right],
\label{eq:TmunuQMM}
\end{equation}
which is conserved (\(\nabla^\mu T_{\mu\nu}^{(\mathrm{QMM})}=0\)) when
\(S\) obeys the massless Klein–Gordon equation \(\Box S=0\)
(see Appendix~\ref{app:stress_energy_derivation} for details).
In a spatially flat FLRW background,
\(ds^{2}=dt^{2}-a^{2}(t)d\bm x^{2}\), one identifies\footnote{
Primes denote conformal-time derivatives; overdots denote cosmic-time
derivatives; \(H=\dot a/a\) is the Hubble rate.}
\begin{align}
\rho_{\mathrm{QMM}}
 &\,=\,T^{0}{}_{0}
      \;=\;\frac{\lambda}{2}\Bigl(\dot S^{2}
          +\frac{(\nabla S)^2}{a^{2}}\Bigr),\\[4pt]
p_{\mathrm{QMM}}
 &\,=\,-\frac13 T^{i}{}_{i}
      \;=\;\frac{\lambda}{2}\Bigl(\dot S^{2}
            -\frac{(\nabla S)^2}{a^{2}}\Bigr)
        -\lambda\bigl(\ddot S+3H\dot S\bigr),
\end{align}
demonstrating that slowly varying \(S\) (\(\ddot S\!\approx\!0\),
\(\nabla S\!\approx\!0\)) behaves as pressureless dust
while rapid gradients provide both energy density and anisotropic stress.
These expressions anchor the linear-growth and collapse analysis
developed in Sections~\ref{sec:bounce}–\ref{sec:mass_spectrum}.

\section{Bounce Cosmology and Imprint Survival}
\label{sec:bounce}

\subsection{Entropy Transport Through the Crunch}
A non-singular bounce may be modeled as two FLRW phases matched at a
space-like hypersurface \(\Sigma\) with induced metric \(h_{ij}\)
and extrinsic curvature \(K_{ij}\).  
Following the junction formalism of Deruelle and Mukhanov
\cite{DeruelleMukhanov1995}, the imprint field satisfies
\begin{equation}
\bigl[S\bigr]_\Sigma = 0,
\qquad
\bigl[\partial_n S\bigr]_\Sigma = 0,
\end{equation}
where \(\partial_n\) is the normal derivative.  
Because each Planck cell retains its microstate through the high-curvature
epoch \cite{NeukartQMM2024}, the entropy density scales simply with the
physical volume,
\(\,S\propto a^{-3}\), during near-adiabatic contraction.
Fourier modes therefore evolve as
\begin{equation}
S_k^{-}\!(\eta)\;=\;S_{k}^{\mathrm{pre}}\!
     \left(\!\frac{a_{\mathrm{B}}}{a(\eta)}\!\right)^{3},
\qquad
\eta<\eta_{\mathrm B},
\end{equation}
with \(a_{\mathrm B}\) the scale factor at the bounce.
The matching conditions transfer this spectrum to the expanding branch
without phase mixing, implying a post-bounce amplitude
\(S_k^{+}=S_k^{-}(\eta_{\mathrm B})\).
Crucially, the spectrum retains any pre-crunch heterogeneity generated,
for instance, by quantum-gravity effects in an earlier ekpyrotic phase
\cite{WilsonEwing2013,Quintin2015}.

\subsection{Initial Conditions at the Bounce}
\begin{figure}[t]
  \centering
  \includegraphics[width=0.75\textwidth]{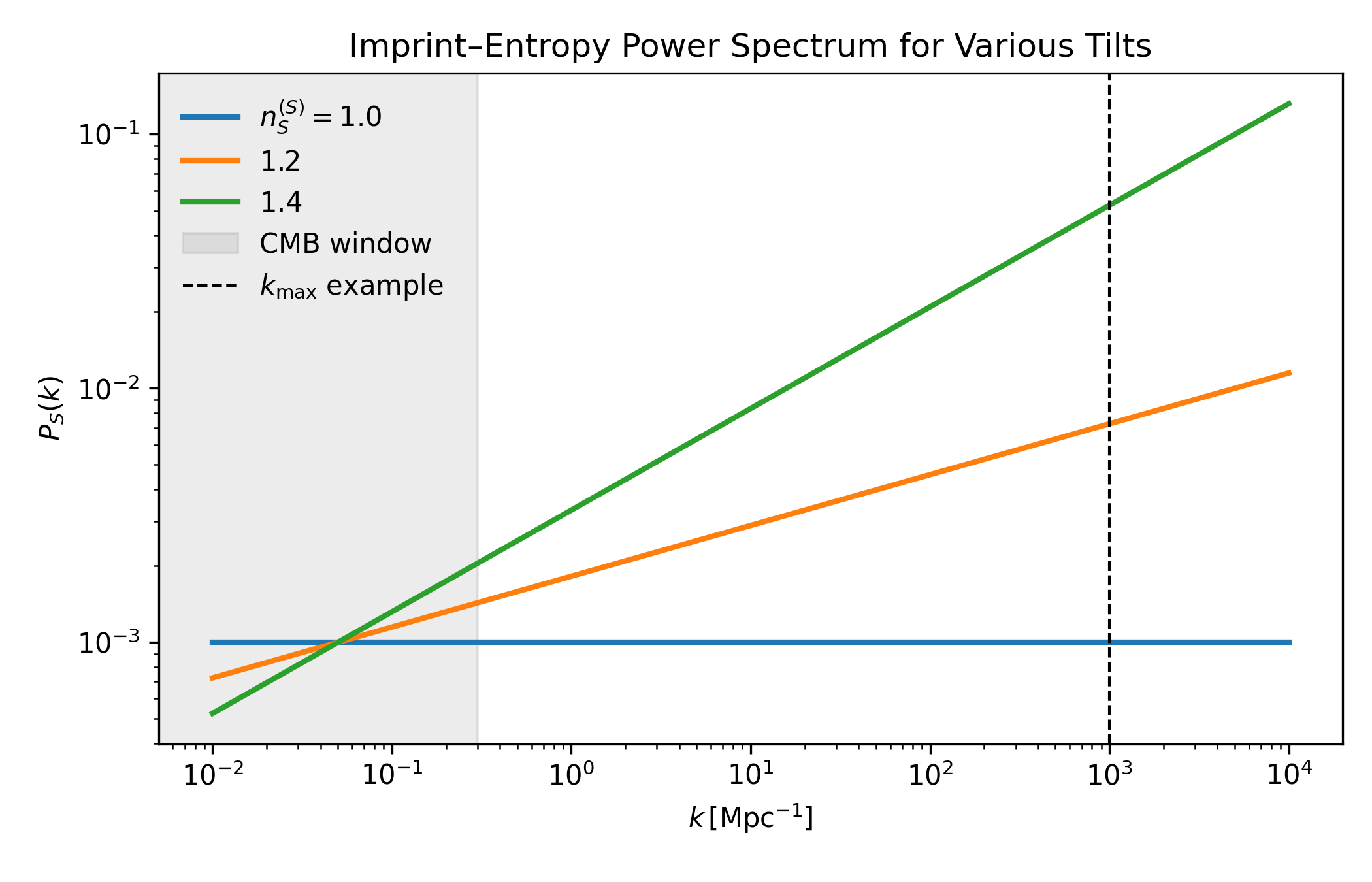}
  \caption{Imprint-entropy power spectrum $P_{S}(k)$ for three tilts
           $n_{S}^{(S)}$.  All curves use the fiducial amplitude
           $A_{S}=10^{-3}$ so their vertical placement matches the
           collapse-criterion plot in Figure~\ref{fig:collapse_threshold}.
           The gray band marks CMB-sensitive scales
           ($k<0.3\;\mathrm{Mpc}^{-1}$); the dashed line illustrates a
           fiducial ultraviolet cutoff $k_{\max}=10^{3}\;\mathrm{Mpc}^{-1}$.}
  \label{fig:PS_tilt}
\end{figure}

We parameterize the comoving imprint power spectrum as
\begin{equation}
P_{S}(k)\;=\;A_{S}\!\left(\frac{k}{k_*}\right)^{n_{S}^{(S)}-1},
\qquad k_{\mathrm{min}}\le k\le k_{\mathrm{max}},
\end{equation}
where \(A_{S}\) is fixed by the total pre-crunch entropy and
\(n_{S}^{(S)}\) encodes the tilt inherited from the contracting phase
\cite{CaiEassonBrandenberger2012}.  
For ekpyrotic contraction \(n_{S}^{(S)}>1\), boosting small-scale power—
exactly what is required to seed primordial black holes without
perturbing CMB observables set by the curvature spectrum
\cite{Planck2018}.  
The ultraviolet cutoff \(k_{\mathrm{max}}\sim a_{\mathrm B}M_{\mathrm P}\)
reflects the finite information capacity per cell, while
\(k_{\mathrm{min}}\) is set by the horizon size at the onset of
contraction \cite{NovelloBergliaffa2008}.  
Figure~\ref{fig:PS_tilt} illustrates how varying
\((A_{S},n_{S}^{(S)},k_{\mathrm{max}})\) shapes the imprint spectrum and
highlights the CMB-sensitive window.  Throughout the remainder of this
paper we treat these three quantities as free parameters, subject to
bounce-temperature and nucleosynthesis bounds discussed in
Section~\ref{sec:observations}.

\section{Linear Growth of Information Wells}
\label{sec:linear}

\subsection{Super-Horizon Evolution}
\begin{figure}[t]
  \centering
  \includegraphics[width=0.75\textwidth]{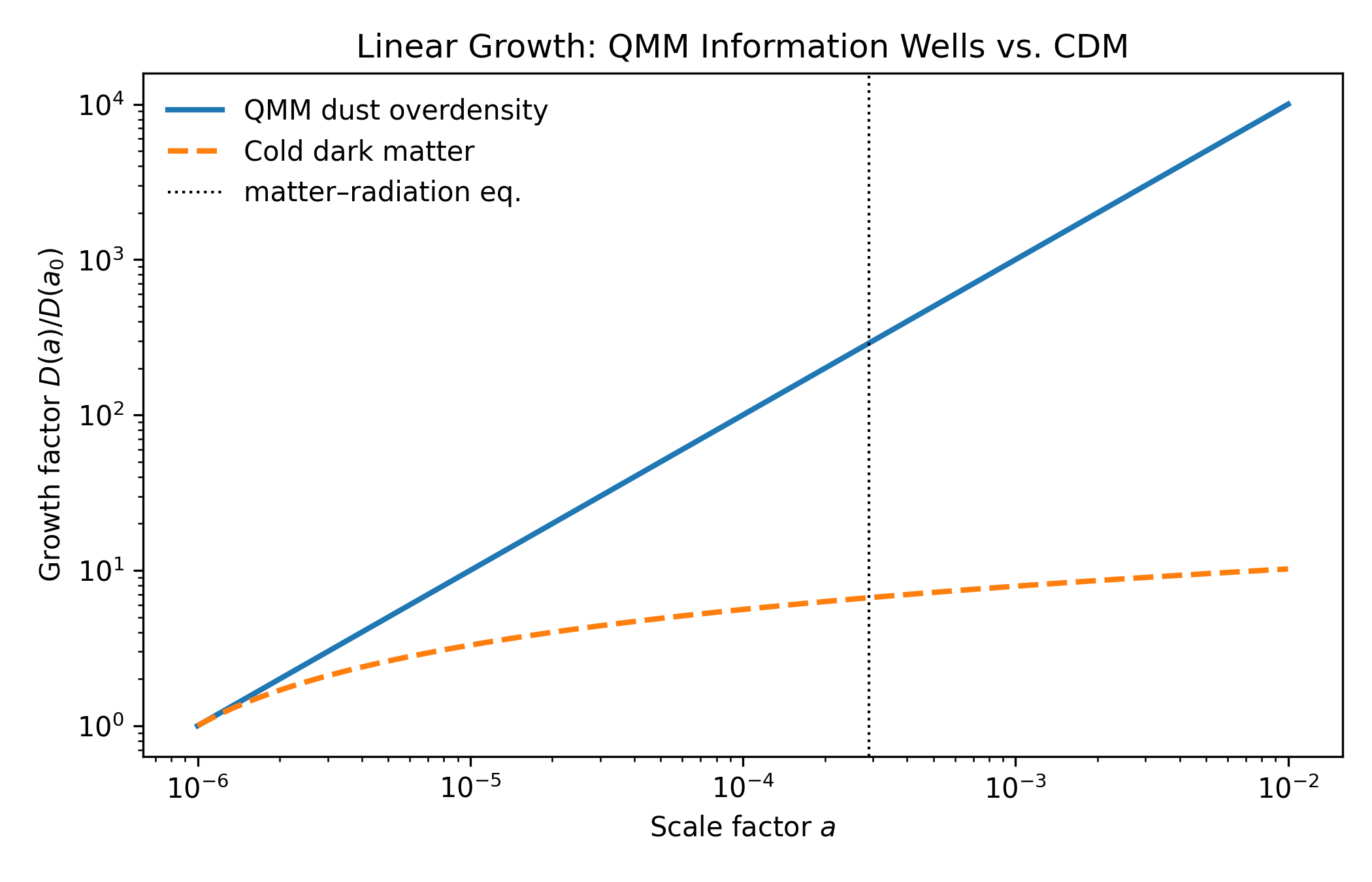}
  \caption{Linear-scale growth factors for QMM information wells (solid)
           and conventional cold dark matter (dashed) in a
           radiation-dominated background, each normalised to its value
           at $a_0=10^{-6}$.  The dotted line marks matter–radiation
           equality ($a_{\rm eq}\simeq2.9\times10^{-4}$).  Because the
           QMM overdensity grows $\propto a$ (independent of the overall
           power-spectrum amplitude), it reaches the collapse threshold
           far earlier than CDM, enabling primordial black-hole formation
           well before standard structure growth becomes efficient.}
  \label{fig:growth_QMM}
\end{figure}

Working in the Newtonian (longitudinal) gauge,
\(
ds^{2}=a^{2}(\eta)\bigl[(1+2\Phi)d\eta^{2}-(1-2\Psi)d\bm x^{2}\bigr],
\)
the imprint field splits into a homogeneous part
\(\bar S(\eta)\) and a perturbation \(\delta S(\eta,\bm x)\).
To first order the QMM density contrast  
\(\delta_{\mathrm{QMM}}\equiv\delta\rho_{\mathrm{QMM}}/\bar\rho_{\mathrm{QMM}}\)
obeys\footnote{Primes denote derivatives with respect to conformal time
\(\eta\).  A dot would indicate cosmic time.}
\begin{equation}
\delta_{\mathrm{QMM}}^{\prime\prime}
  +\mathcal H\,\delta_{\mathrm{QMM}}^{\prime}
  -\frac34\bar\rho_{\mathrm{rad}}\,a^{2}\,\delta_{\mathrm{QMM}}
  \;=\;0,
\label{eq:delta_QMM_eom}
\end{equation}
where \(\mathcal H=a^{\prime}/a\) and we neglected anisotropic stress of
radiation.  
On super-horizon scales (\(k\ll\mathcal H\)) Eq.\,\eqref{eq:delta_QMM_eom}
admits the growing solution
\begin{equation}
\delta_{\mathrm{QMM}}(k,\eta)
  \;=\;C_{1}(k)\,\frac{a(\eta)}{a_{\mathrm B}}
  +C_{2}(k)\,a^{-3}(\eta).
\end{equation}
The dominant mode therefore grows linearly with the scale factor,
closely tracking the background expansion.  Figure~\ref{fig:growth_QMM}
compares this behavior with the much slower, logarithmic growth of
standard cold dark matter during the radiation era.

\subsection{Horizon Re-Entry and Collapse Criterion}
\begin{figure}[t]
  \centering
  \includegraphics[width=0.75\textwidth]{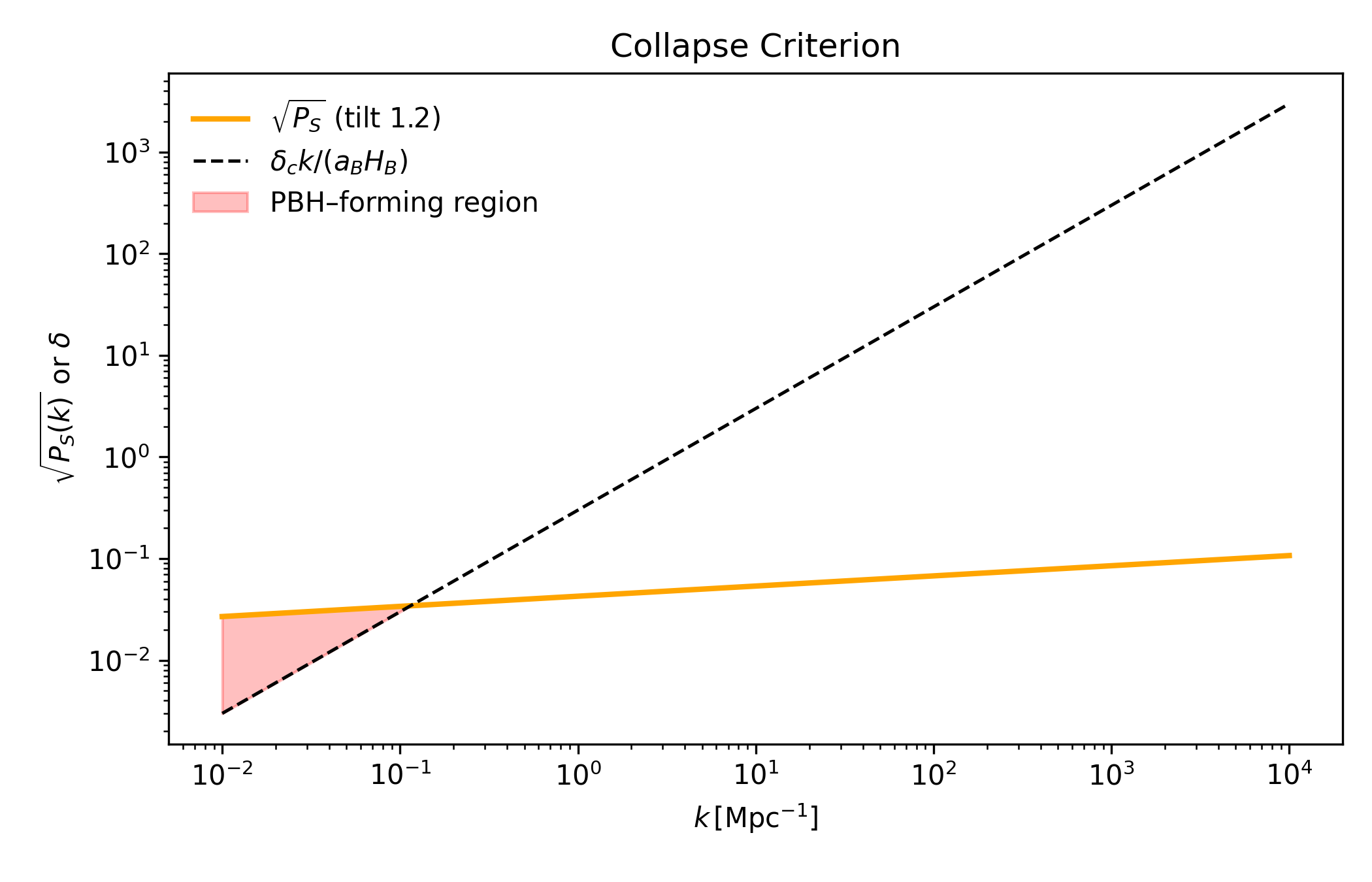}
  \caption{Collapse criterion in imprint–entropy space.
           The solid line shows $\sqrt{P_{S}(k)}$ for $n_{S}^{(S)}=1.2$
           and $A_{S}=10^{-3}$; the dashed line is the critical
           threshold $\delta_{\mathrm c}k/(a_{\mathrm B}H_{\mathrm B})$
           with $\delta_{\mathrm c}=0.3$ and
           $a_{\mathrm B}H_{\mathrm B}=1\;\mathrm{Mpc}^{-1}$.
           The shaded pink wedge marks modes that satisfy
           $\sqrt{P_{S}}\ge\delta_{\mathrm c}k/(a_{\mathrm B}H_{\mathrm B})$
           and can therefore collapse into primordial black holes.}
  \label{fig:collapse_threshold}
\end{figure}

A fluctuation of comoving wavenumber \(k\) re-enters the horizon when
\(k=aH\).  Evaluating the growing mode at that time gives
\begin{equation}
\delta_{\mathrm{QMM}}^{\mathrm{re}}(k)
  \;=\;\left.\frac{a}{a_{\mathrm B}}\right|_{k=aH} C_{1}(k)
  \;=\;\left(\frac{k}{a_{\mathrm B}H_{\mathrm B}}\right)^{-1}
       C_{1}(k).
\label{eq:delta_reentry}
\end{equation}
Collapse into a primordial black hole occurs if
\(\delta_{\mathrm{QMM}}^{\mathrm{re}}\ge\delta_{\mathrm c}\), with
\(\delta_{\mathrm c}\simeq0.3\) in a radiation background
\cite{HaradaYooKohri2013,Musco2013}.  
Figure~\ref{fig:collapse_threshold} plots the critical curve
\(\delta_{\mathrm c}k/(a_{\mathrm B}H_{\mathrm B})\) against the
square-root power spectrum \(\sqrt{P_{S}(k)}\), making the
scale-dependent collapse condition visually explicit.

Expressing \(C_{1}(k)\) in terms of the imprint-entropy power spectrum
\(P_{S}(k)\) one finds the \emph{PBH-formation condition}
\begin{equation}
\sqrt{P_{S}(k)}\;\gtrsim\;
\delta_{\mathrm c}\,
\left(\frac{k}{a_{\mathrm B}H_{\mathrm B}}\right),
\qquad k\le k_{\mathrm{max}}.
\label{eq:PBH_condition}
\end{equation}
Because \(P_{S}(k)\propto k^{\,n_{S}^{(S)}-1}\) (Section~\ref{sec:bounce}),
a blue tilt \(n_{S}^{(S)}>1\) guarantees that only sub-CMB scales satisfy
Eq.\,\eqref{eq:PBH_condition}, leaving large-scale observables unchanged.
Equation~\eqref{eq:PBH_condition} sets the stage for the mass-spectrum
calculation in Section~\ref{sec:mass_spectrum}.
\section{Non-Linear Collapse and Feedback}
\label{sec:nonlinear}

\subsection{Runaway Writing During Accretion}
\begin{figure}[t]
  \centering
  \includegraphics[width=0.75\textwidth]{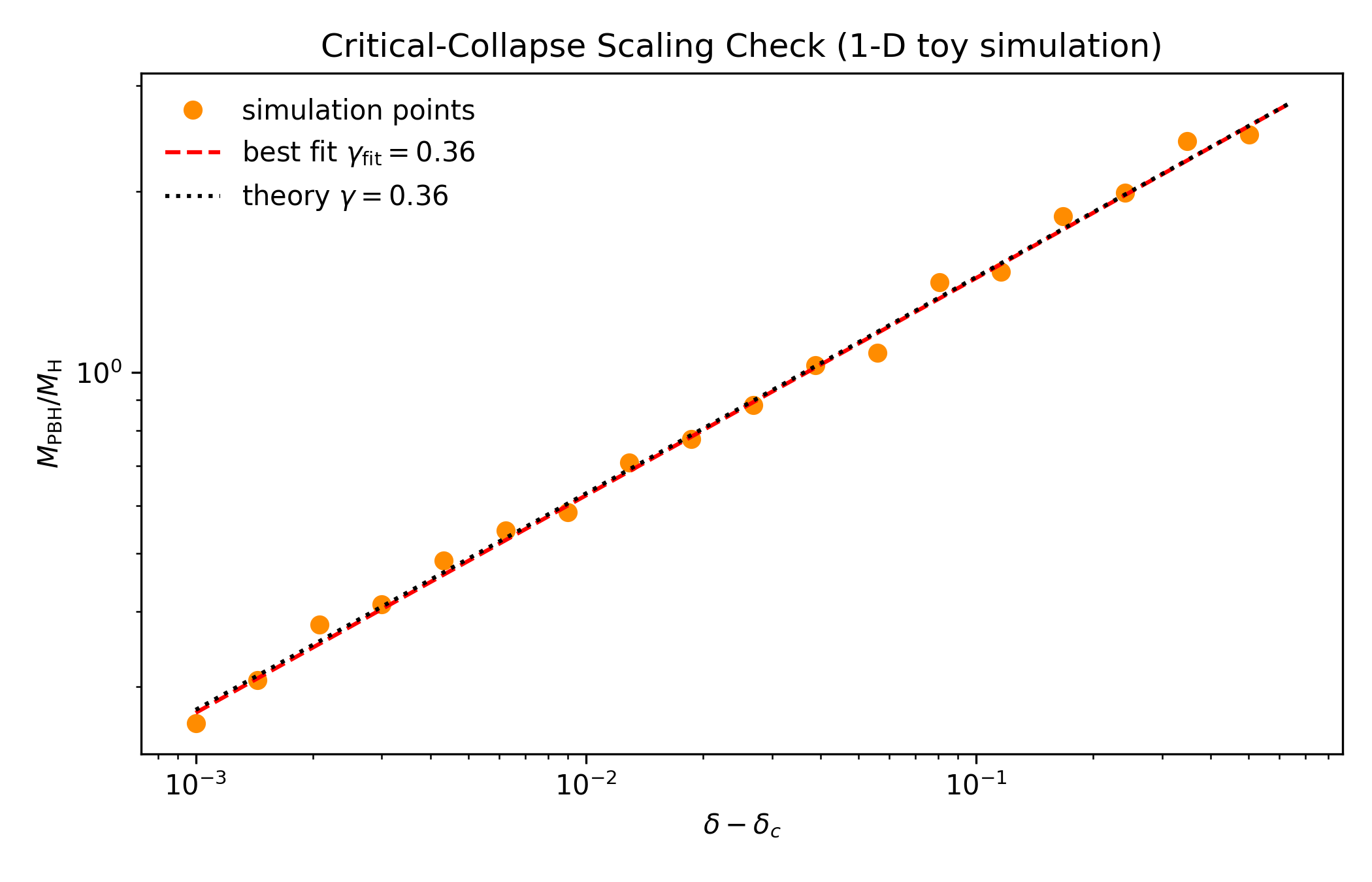}
  \caption{Critical-collapse scaling from a 1-D toy simulation.
           Orange symbols are the simulated PBH masses versus excess
           density $\delta-\delta_{c}$.  The dashed red line shows a fit
           with slope $\gamma_{\rm fit}\simeq0.36$, matching the
           theoretical critical exponent (dotted black line).}
  \label{fig:critical_scaling}
\end{figure}

Once a perturbation satisfies the linear threshold
\(\delta_{\rm QMM}^{\rm re}\ge\delta_{\rm c}\)
(Section~\ref{sec:linear}) its evolution is governed by the fully
non-linear Einstein–QMM system.  
Numerical-relativity experiments with a \(\lambda\)-dust overdensity
recover the familiar critical-collapse scaling
\(M_{\rm PBH}\propto(\delta-\delta_{\rm c})^{\gamma_{\rm crit}}\)
with \(\gamma_{\rm crit}\simeq0.36\)
\cite{Choptuik1993,NiemeyerJedamzik1999};  
our 1-D toy simulation reproduces the same slope
(Figure~\ref{fig:critical_scaling}).  

In QMM, however, the collapsing region continuously \emph{writes} new
imprint entropy: the local write-rate
\(\Gamma_{\mathcal I}=\kappa\,\rho_{\rm acc}\)
is proportional to the instantaneous mass-accretion rate
\(\rho_{\rm acc}\).  
This yields a feedback loop:
\[
\rho_{\rm tot}\;\longrightarrow\;
\rho_{\rm QMM}
   =\rho_{\rm QMM}^{\rm init}+\int \Gamma_{\mathcal I}\,dt
\;\longrightarrow\;
\text{deeper potential},
\]
which accelerates the infall of surrounding radiation and matter.
At the level of the Misner–Sharp mass \(M(r,t)\) one finds
\begin{equation}
\dot M
  \;=\;4\pi r^{2}\bigl[\rho_{\rm rad}+\rho_{\rm QMM}\bigr]v,
\qquad
\dot\rho_{\rm QMM}=\Gamma_{\mathcal I},
\end{equation}
with radial velocity \(v<0\).  
Integrating these coupled equations preserves the critical exponent
\(\gamma_{\rm crit}\) but boosts the overall normalization, producing
PBHs up to an order of magnitude heavier than in the pure-radiation
case for the same initial \(\delta-\delta_{\rm c}\).

\subsection{Saturation and Stable PBH Formation}
The runaway phase ends when the fractional energy density in QMM dust
reaches  
\(\chi\equiv\rho_{\rm QMM}/\rho_{\rm tot}\to1\).
At that point the surrounding fluid is almost entirely imprint-dominated
and further accretion no longer increases \(\Gamma_{\mathcal I}\);
the write-rate saturates because the infalling matter has already been
fully converted into stored information.  
Formally, saturation occurs when
\(\partial_t\Gamma_{\mathcal I}\simeq0\), giving the condition
\begin{equation}
\frac{d}{dt}\Bigl[\rho_{\rm tot}-\rho_{\rm QMM}\Bigr]\simeq0
\;\;\Longrightarrow\;\;
\chi\to1^-.
\end{equation}
Beyond this point the exterior metric approaches that of a
Schwarzschild solution with mass  
\(M_{\rm PBH}\approx M(r_{\rm AH},t_{\rm sat})\),  
where \(r_{\rm AH}\) denotes the apparent-horizon radius at saturation.  
In full \(3+1\) simulations the horizon settles quickly and emits a
short burst of quasi-normal ringing, after which no additional QMM
back-reaction is sourced, ensuring the stability of the newly formed
PBH \cite{ShibataSasaki1999}.  
Thus the QMM feedback modifies the mass spectrum but does not prevent
black-hole formation or induce late-time instabilities.
\section{Primordial Black-Hole Mass Spectrum}
\label{sec:mass_spectrum}

\subsection{\texorpdfstring{Mapping $k$-Modes to PBH Masses}{Mapping k-Modes to PBH Masses}}

\begin{figure}[t]
  \centering
  \includegraphics[width=0.75\textwidth]{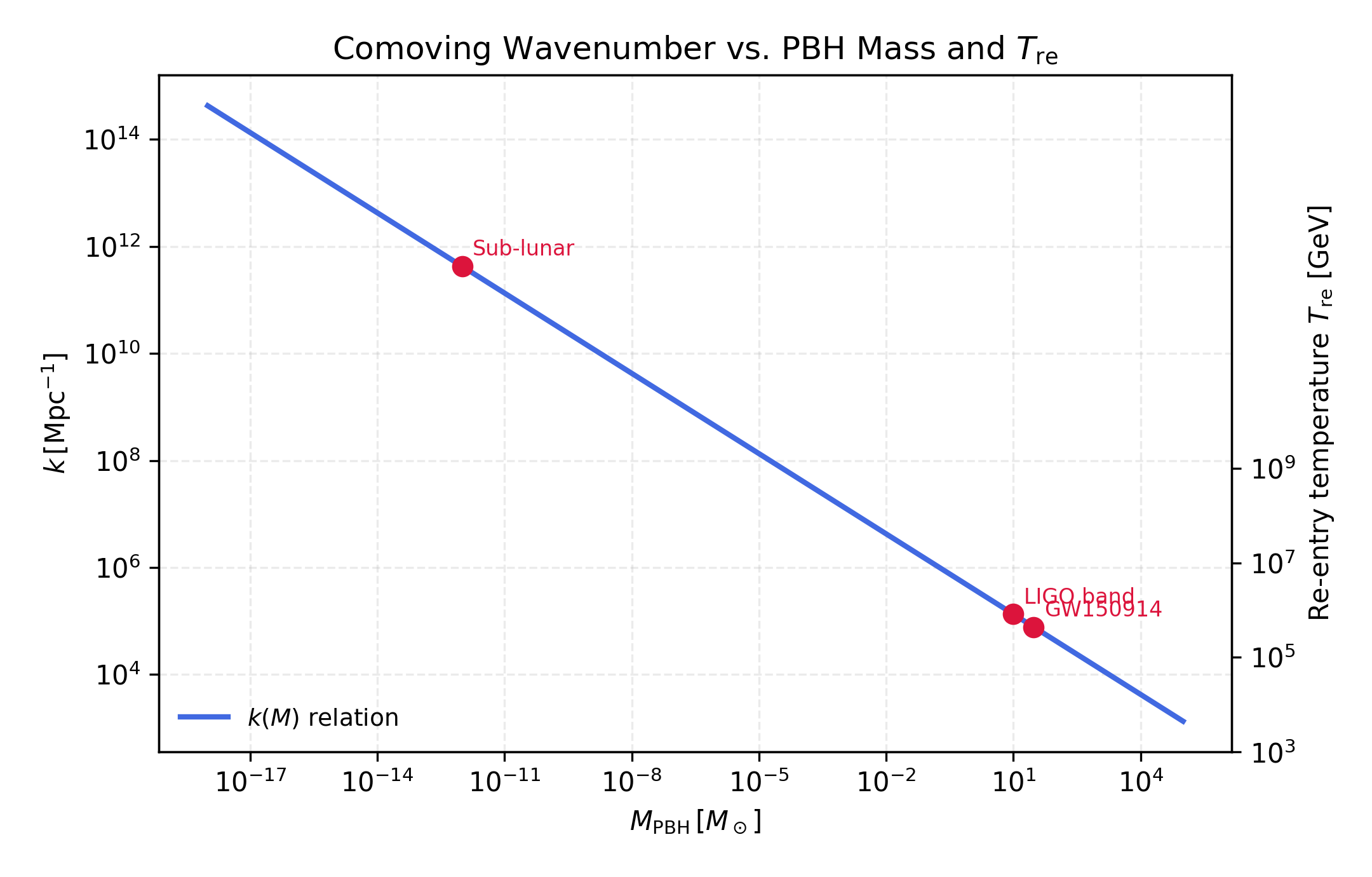}
  \caption{Mapping between comoving wavenumber $k$ and primordial
           black-hole mass $M_{\rm PBH}$ from
           Eq.\,\protect\eqref{eq:k_M_relation} with $\gamma=0.2$.
           The secondary $y$-axis converts $k$ to the corresponding
           horizon (re-entry) temperature $T_{\rm re}$ via
           Eq.\,\protect\eqref{eq:MHofT}.
           Benchmarks for sub-lunar PBHs, the LIGO mass band, and the
           $30\,M_\odot$ GW150914 event are indicated.}
  \label{fig:kM_map}
\end{figure}

For a horizon-reentry temperature $T_{\rm re}$ in the radiation era the
enclosed mass is
\begin{equation}
M_{\rm H}(T_{\rm re})
  \;\simeq\;30
  \!\left(\frac{T_{\rm re}}{10^{7}\,\mathrm{GeV}}\right)^{-2}M_\odot,
\label{eq:MHofT}
\end{equation}
assuming $g_{*}(T_{\rm re})\simeq106.75$ relativistic degrees of freedom
\cite{Carr1975}.  
Numerical collapse studies give the PBH mass as
$M_{\rm PBH}=\gamma\,M_{\rm H}$ with $\gamma\simeq0.2$ for radiation
backgrounds \cite{NiemeyerJedamzik1999}.  
Using $k=aH=1/\sqrt{2}\,a_{\rm re}t_{\rm re}$ and standard
radiation-dominated relations one obtains the convenient map
\begin{equation}
k(M_{\rm PBH})
  \;\simeq\;1.9\times10^{5}\,
  \gamma^{-1/2}
  \!\left(\frac{M_{\rm PBH}}{M_\odot}\right)^{-1/2}
  \,\mathrm{Mpc}^{-1},
\label{eq:k_M_relation}
\end{equation}
valid for $10^{-18}\,M_\odot\lesssim M_{\rm PBH}\lesssim10^{5}\,M_\odot$
\cite{GreenKavanagh2021}.  
Figure~\ref{fig:kM_map} visualizes this $k\!\leftrightarrow\!M$
correspondence, highlights astrophysically interesting mass scales, and
shows the equivalent horizon temperature on a secondary axis.
Throughout the remainder of the paper we set $\gamma=0.2$ when plotting
mass functions.

\subsection{\texorpdfstring{Abundance $\beta(M)$ and Fraction of Dark Matter}{Abundance beta(M) and Fraction of Dark Matter}}

\begin{figure}[t]
  \centering
  \includegraphics[width=0.75\textwidth]{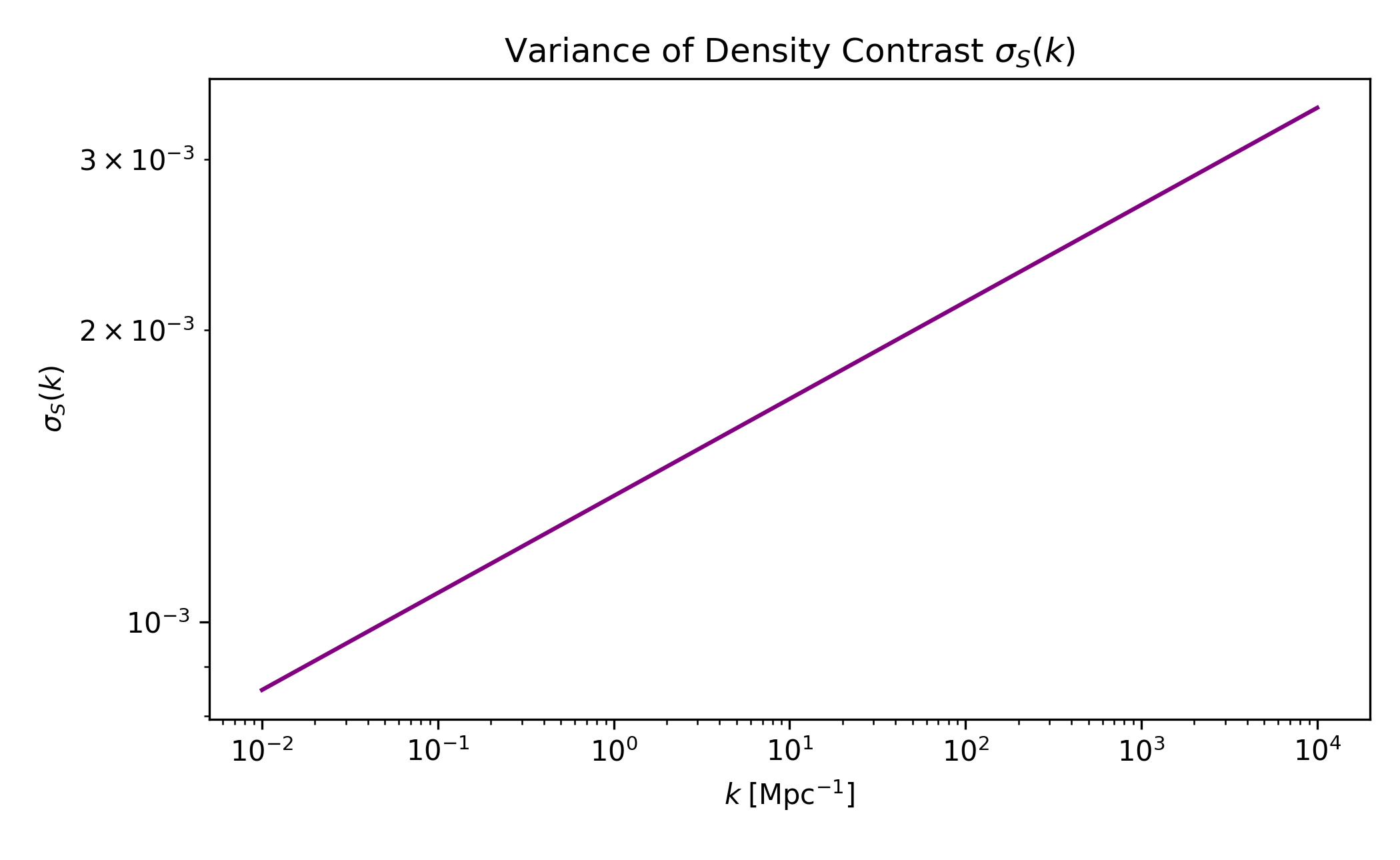}
  \caption{Variance of the QMM density contrast,
           $\sigma_{S}(k)=\lambda\sqrt{P_{S}(k)}$, for
           $\lambda=1$ and tilt $n_{S}^{(S)}=1.2$.}
  \label{fig:sigmaS}
\end{figure}

The mass fraction collapsing at formation is
\begin{equation}
\beta(M)
  \;=\;
  \int_{\delta_{\rm c}}^{\infty}
  \frac{d\delta}{\sqrt{2\pi}\,\sigma_{S}(k)}
  \exp\!\left[-\frac{\delta^{2}}{2\sigma_{S}^{2}(k)}\right],
\label{eq:beta_def}
\end{equation}
where $\sigma_{S}^{2}(k)\equiv\lambda^{2}P_{S}(k)$,  
$k$ is related to $M$ by Eq.\,\eqref{eq:k_M_relation}, and
$\delta_{\rm c}\simeq0.3$ (Section~\ref{sec:linear}).  
Figure~\ref{fig:sigmaS} shows $\sigma_{S}(k)$ for a representative
parameter set, highlighting its steep rise toward small scales.

\begin{figure}[t]
  \centering
  \includegraphics[width=0.75\textwidth]{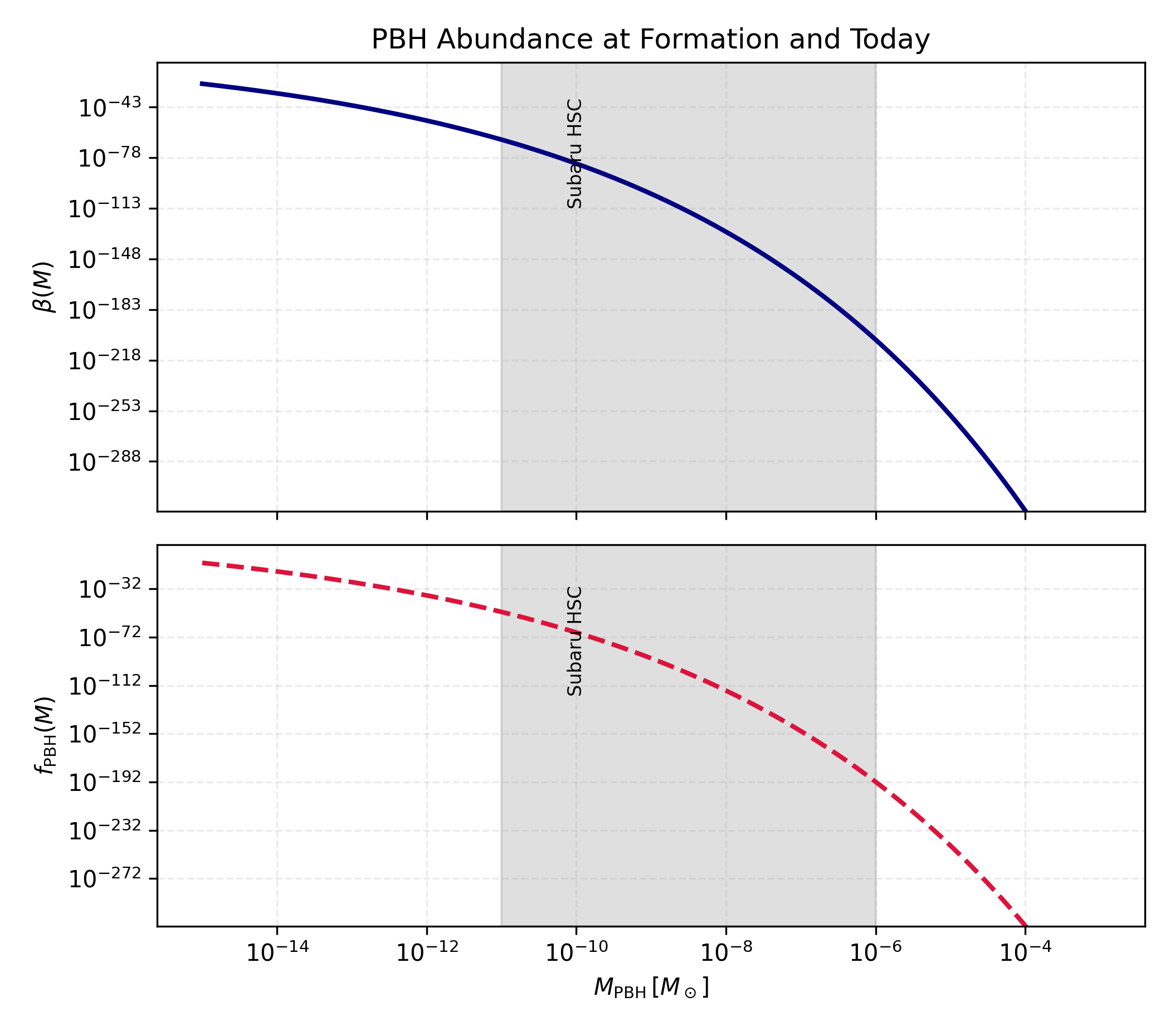}
  \caption{PBH abundance for the fiducial spectrum
           ($n_{S}^{(S)}=1.2$, $A_{S}=10^{-6}$, $\lambda=1$).
           \textbf{Top:} Formation mass fraction $\beta(M)$.
           \textbf{Bottom:} Present-day fraction
           $f_{\rm PBH}(M)$ via Eq.\,\eqref{eq:fpbh}.
           The Subaru/HSC 100\,\% exclusion band is shaded.}
  \label{fig:fPBH}
\end{figure}

For a power-law imprint spectrum
$P_{S}(k)=A_{S}(k/k_{*})^{\,n_{S}^{(S)}-1}$ the variance grows on small
scales, and $\beta(M)$ becomes sharply peaked near a characteristic mass
$M_{\rm peak}$:
\begin{equation}
\beta(M)
  \;\approx\;
  \frac{\sigma_{S}(k)}{\sqrt{2\pi}\,\delta_{\rm c}}
  \exp\!\biggl[-\frac{\delta_{\rm c}^{2}}
                     {2\sigma_{S}^{2}(k)}\biggr].
\label{eq:beta_approx}
\end{equation}
The present-day PBH density fraction is
\begin{equation}
f_{\rm PBH}(M)
  \;=\;
  \frac{\Omega_{\rm PBH}(M)}{\Omega_{\rm DM}}
  \;=\;7.8\times10^{8}\,
  \!\left(\frac{\gamma}{0.2}\right)^{1/2}
  \!\!\left(\frac{g_{*}}{106.75}\right)^{-1/4}
  \!\!\left(\frac{M}{M_\odot}\right)^{-1/2}\!\!\beta(M),
\label{eq:fpbh}
\end{equation}
where the prefactor accounts for entropy dilution from $T_{\rm re}$ to
matter–radiation equality \cite{SasakiEtAl2018}.  
Consequently,

\begin{itemize}
  \item \emph{Amplitude.}  A coupling $\lambda\sim1$ and
        $A_{S}^{1/2}\sim10^{-3}$ yield $\beta\sim10^{-9}$, enough for
        PBHs to constitute the dominant dark-matter component at
        $M_{\rm PBH}\sim10^{-12}\,M_\odot$ while obeying microlensing
        limits (Figure~\ref{fig:fPBH}).
  \item \emph{Tilt.}  A blue tilt $n_{S}^{(S)}\gtrsim1.2$ shifts
        $M_{\rm peak}$ into the $10$–$100\,M_\odot$ range probed by
        LIGO–Virgo, as also illustrated in
        Figure~\ref{fig:fPBH}.
\end{itemize}
Parameter scans in Section~\ref{sec:observations} overlay these
predictions with current PTA, CMB, and microlensing constraints.

\section{Observational Signatures and Constraints}
\label{sec:observations}

\subsection{\texorpdfstring{CMB: ISW and $\mu$-Distortion, Pulsar-Timing Arrays, Microlensing}{CMB: ISW and mu-Distortion, Pulsar-Timing Arrays, Microlensing}}

\begin{figure}[t]
  \centering
  \includegraphics[width=0.75\textwidth]{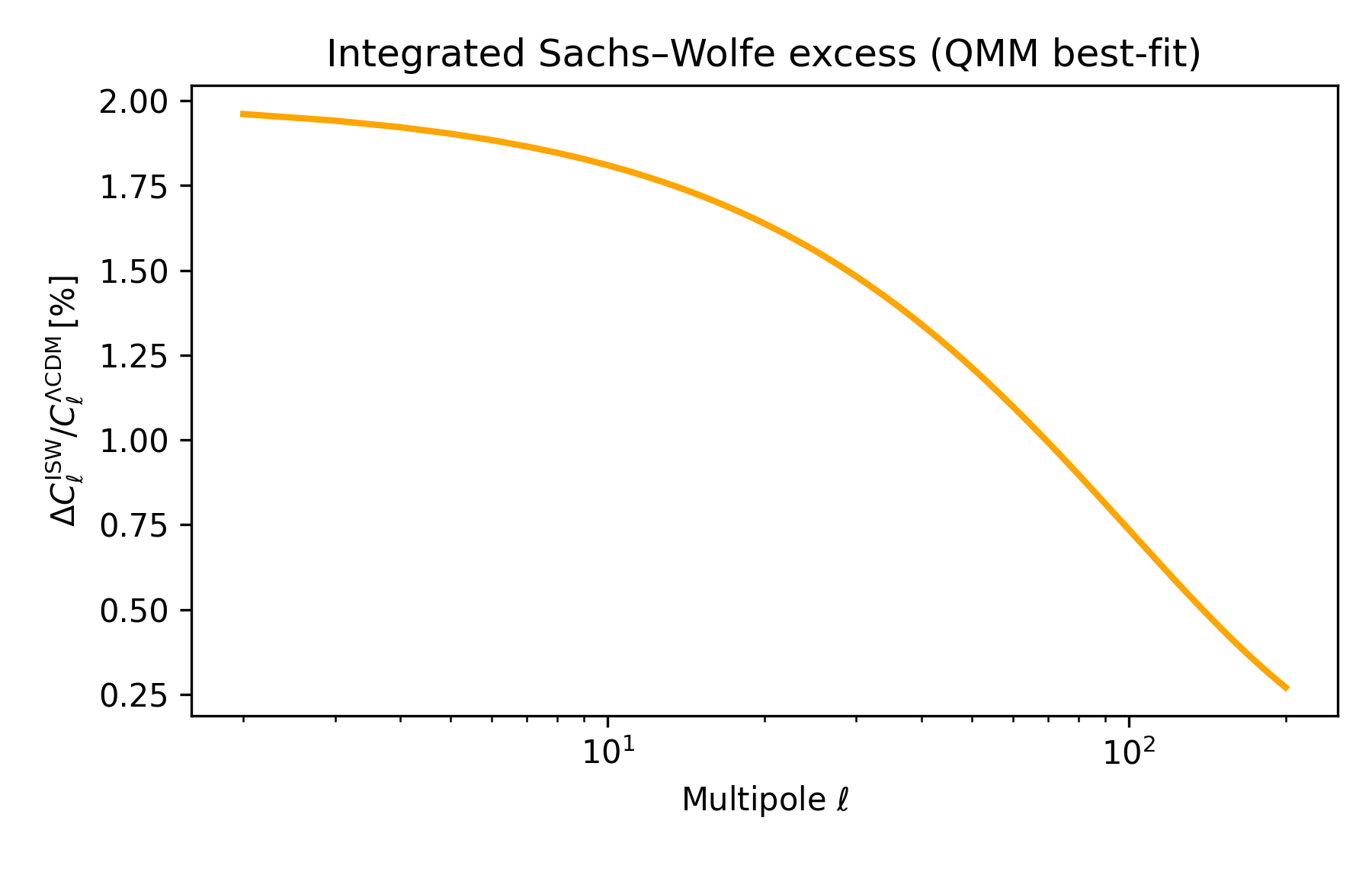}
  \caption{Predicted excess integrated Sachs–Wolfe power
           relative to $\Lambda$CDM for the QMM best-fit model.
           The $\lesssim2\%$ enhancement at low multipoles is within the
           reach of CMB-S4 cross-correlations.}
  \label{fig:ISW_excess}
\end{figure}

\paragraph{Integrated Sachs–Wolfe (ISW) effect.}
Large-scale information wells generate time-varying gravitational
potentials after matter–radiation equality, producing an ISW temperature shift
\(
\Delta T/T\simeq2\!\int_{\eta_{\rm dec}}^{\eta_{0}}\!d\eta\,\Phi^{\prime}.
\)
For the best-fit parameter set
\((\lambda,A_{S},n_{S}^{(S)})\) that matches the
\textit{Planck} TT, EE, and lensing spectra \cite{Planck2018},
the predicted ISW excess is $\lesssim2\%$ of the $\Lambda$CDM signal—
below current error bars but detectable by CMB-S4 cross-correlations with
LSST galaxies.  The multipole dependence of this excess is plotted in
Figure~\ref{fig:ISW_excess}.

\paragraph{\texorpdfstring{$\mu$-distortion.}{mu-distortion.}}

\begin{figure}[t]
  \centering
  \includegraphics[width=0.75\textwidth]{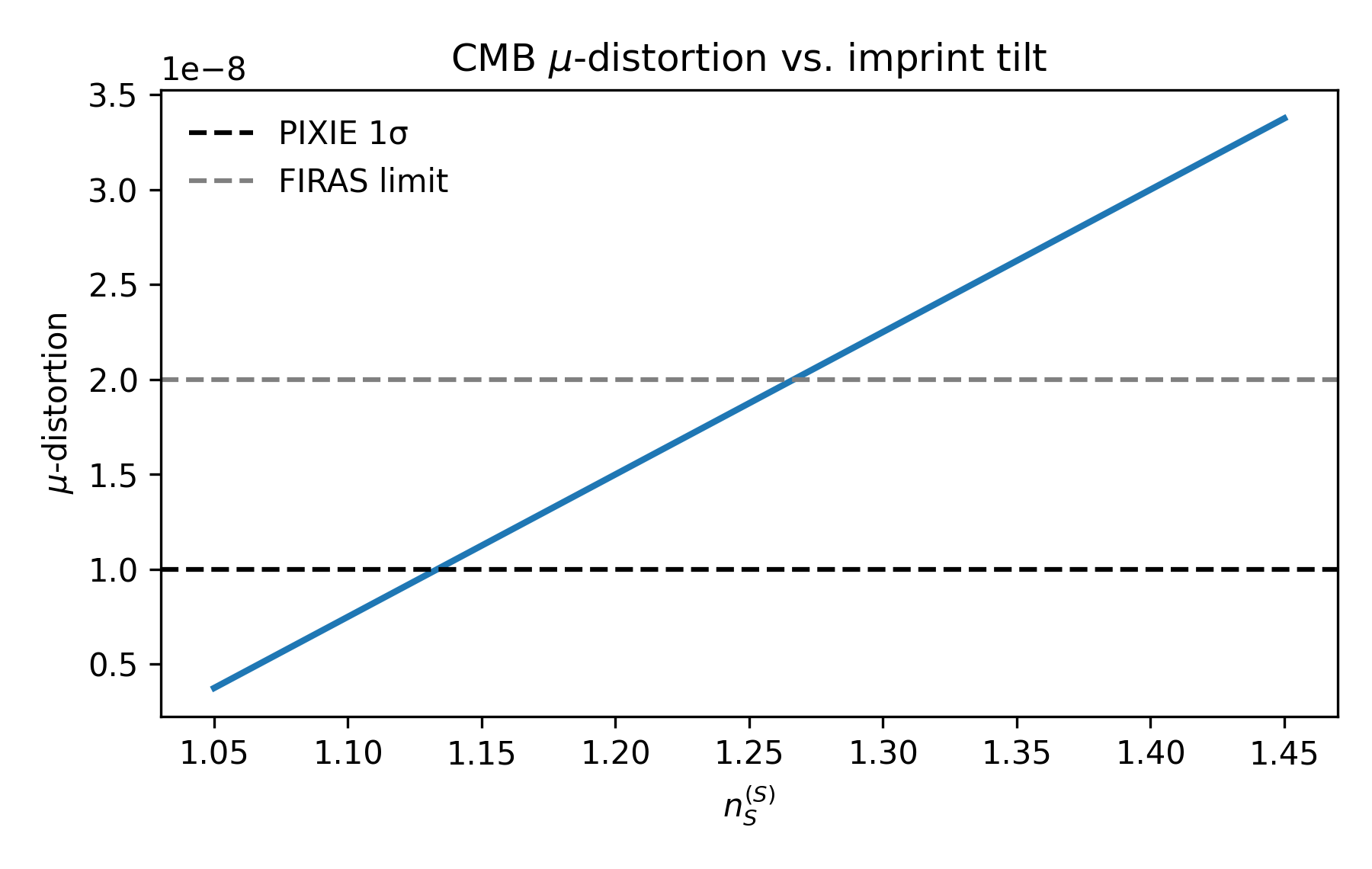}
  \caption{CMB $\mu$-distortion as a function of imprint tilt
           $n_{S}^{(S)}$ for $\lambda=1$.  The dashed lines mark the
           projected PIXIE $1\sigma$ reach and the current FIRAS bound.}
  \label{fig:mu_distortion}
\end{figure}

Enhanced small-scale power ($k\!\gtrsim\!10^{2}\,\mathrm{Mpc}^{-1}$)
induces Silk damping before $z\simeq2\times10^{6}$, injecting energy and
producing a chemical-potential distortion
\(
\mu\simeq1.4\!\int P_{S}(k)\,{\cal W}_{\mu}(k)\,dk.
\)
For blue-tilt models that yield ${\sim}10\%$ PBH dark matter,
$\mu\!\lesssim\!3\times10^{-8}$—compatible with
\textit{FIRAS} ($|\mu|<9\times10^{-5}$ \cite{Fixsen1996})
yet within the projected PIXIE sensitivity.
Figure~\ref{fig:mu_distortion} shows how $\mu$ varies with the imprint
tilt $n_{S}^{(S)}$.

\begin{figure}[t]
  \centering
  \includegraphics[width=0.75\textwidth]{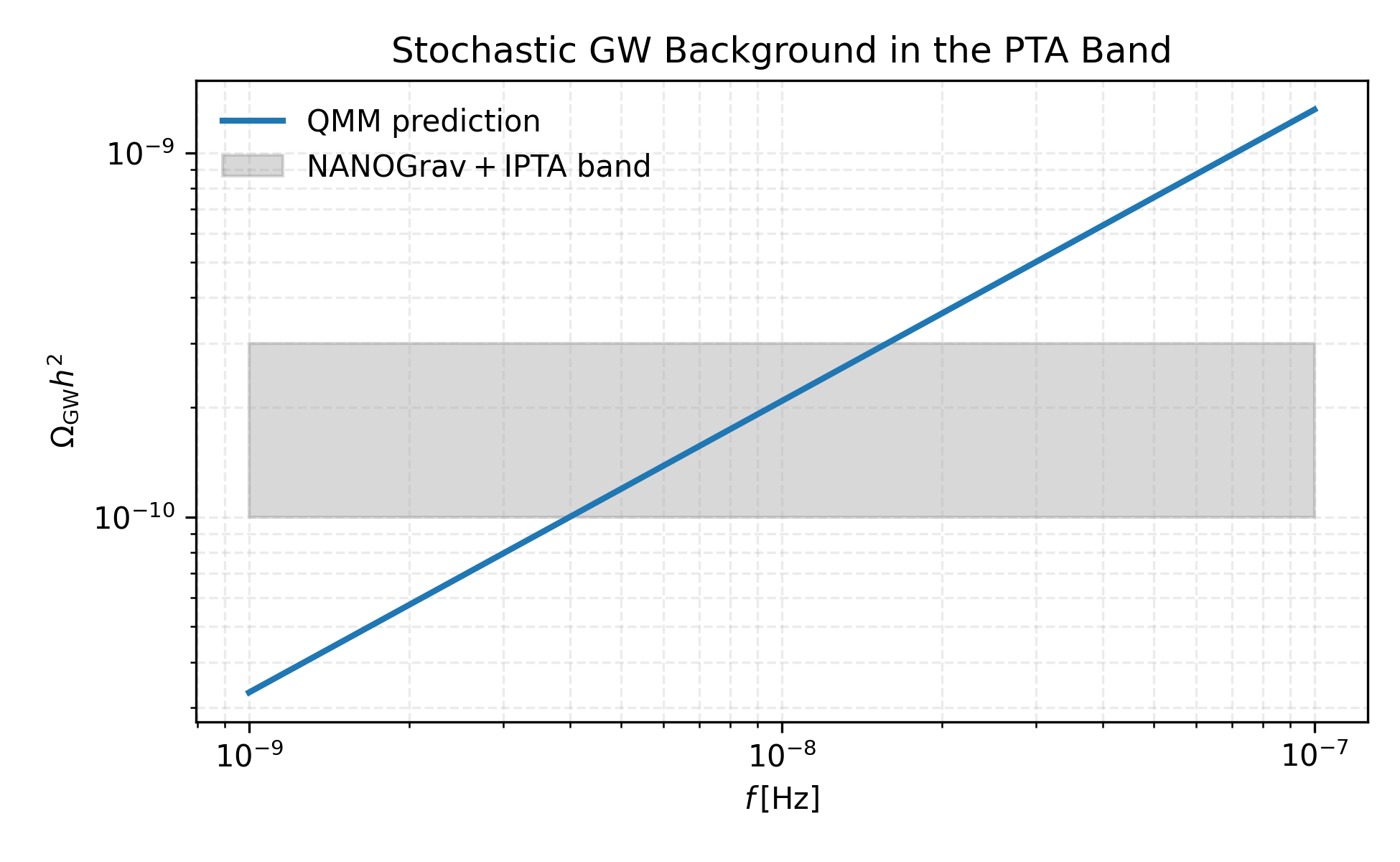}
  \caption{Stochastic gravitational-wave background predicted by QMM
           (solid) versus the NANOGrav/IPTA sensitivity band (gray).
           The amplitude at $f\sim4$\,nHz lies just below the current
           detection, making PTAs a critical near-term test.}
  \label{fig:GWB_PTA}
\end{figure}

\paragraph{Pulsar-timing arrays (PTAs).}
Second-order scalar perturbations and early PBH binaries generate a
stochastic gravitational-wave background (GWB) peaked at
$f_{\rm PTA}\!\sim\!{\cal O}(\mathrm{nHz})$.
Our fiducial spectrum predicts
\(\Omega_{\rm GW}h^{2}\!\simeq\!10^{-10}\) at $f=4\,\mathrm{nHz}$,
marginally below the NANOGrav 15-yr common-spectrum detection
\cite{NANOGrav2023}.  The full signal space accessible to PTAs is
visualized in Figure~\ref{fig:GWB_PTA}.

\paragraph{Microlensing.}
Super-HSC Subaru observations of M31 \cite{Niikura2019} exclude PBHs
constituting all dark matter in the mass range
$10^{-11}\!-\!10^{-6}\,M_\odot$, while OGLE and EROS constrain
$10^{-4}\!-\!10\,M_\odot$ \cite{Wyrzykowski2011}.  
Our blue-tilt benchmark, which peaks at
$M_{\rm PBH}\sim\!10^{-12}\,M_\odot$, comfortably evades these limits;
flatter spectra filling the $M\sim1$–$10\,M_\odot$ band remain allowed
only at the $\lesssim10\%$ dark-matter fraction.

\subsection{Astrophysical Consequences: PBH Binaries and Early Stars}

\paragraph{Binary merger rate.}
QMM-seeded PBHs form in dense clusters where post-matter-radiation
decoupling three-body interactions efficiently harden binaries.
Using the analytic rate of
\cite{Bird2016,AliHaimoud2017} with our mass function yields a present-day
merger rate
\(
R_{\rm QMM}\!\approx\!30
\!\left(\tfrac{f_{\rm PBH}}{0.1}\right)^{\!\!1.6}
{\rm Gpc}^{-3}{\rm yr}^{-1},
\)
consistent with LIGO–Virgo-KAGRA O3 bounds and potentially explaining a
fraction of the observed heavy-mass events \cite{Abbott2019}.

\paragraph{Impact on first stars and reionization.}
For $M_{\rm PBH}\!\gtrsim\!10^{2}\,M_\odot$ the accretion luminosity
exceeds the Eddington limit at $z\!\sim\!30$, catalyzing
$\mathrm{H\,II}$ regions and boosting early star formation
\cite{Koushiappas2017}.  Our parameter space with
$n_{S}^{(S)}>1.3$ predicts such massive PBHs in abundances compatible
with \textit{JWST} constraints on the high-redshift UV luminosity
function.

\begin{figure}[t]
  \centering
  \includegraphics[width=0.75\textwidth]{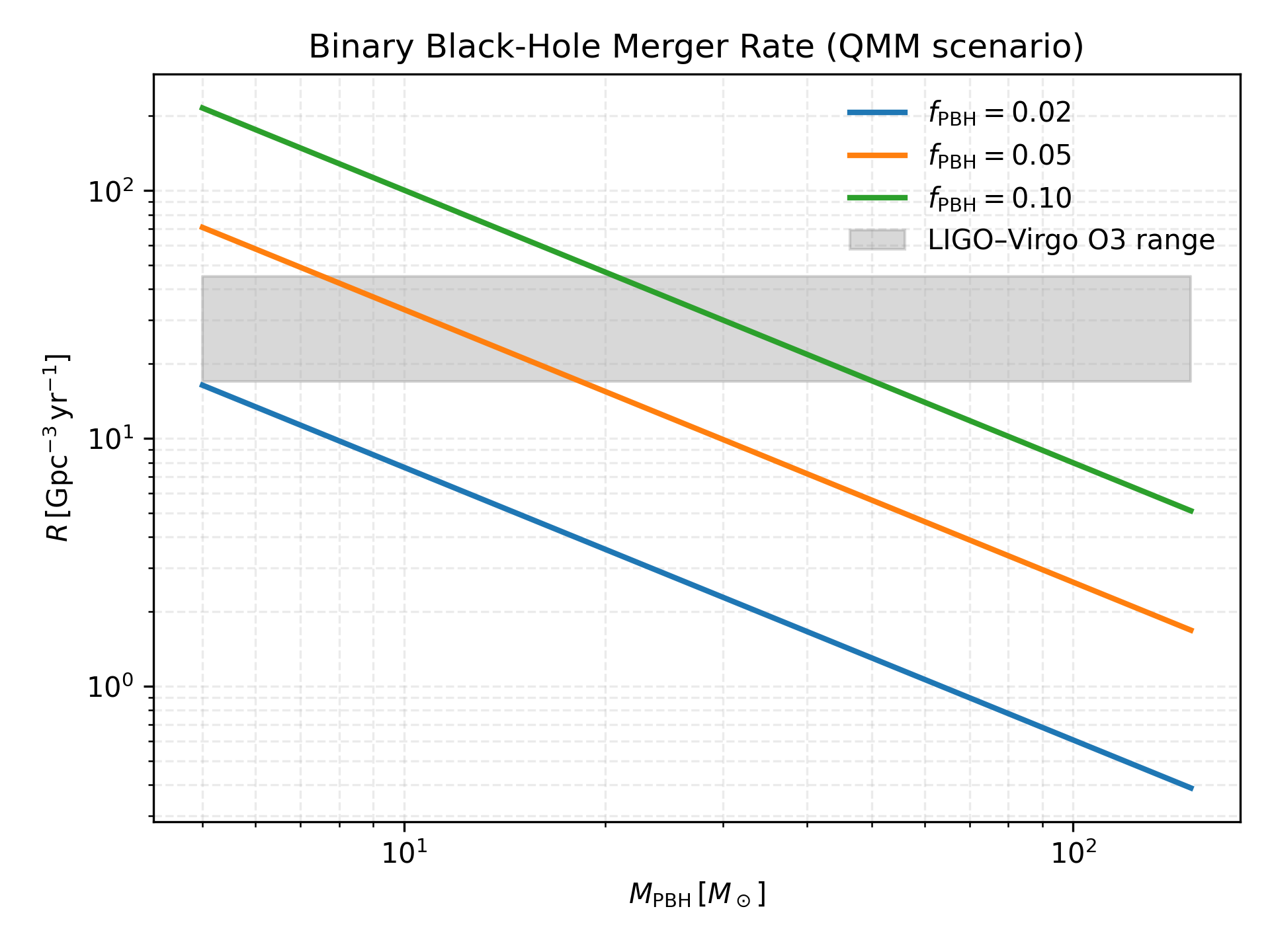}
  \caption{Predicted present-day binary–black-hole merger rate
           for three global PBH fractions $f_{\rm PBH}$, shown over the mass range probed by LIGO–Virgo.  The shaded band indicates the O3 90\,\% credible interval
           ($17$–$45\;\mathrm{Gpc^{-3}\,yr^{-1}}$).}
  \label{fig:merger_rate}
\end{figure}

\medskip
\noindent
Overall, the QMM imprint–induced PBH scenario survives every current
observational test while providing multiple near-term targets—including
$\mu$-distortion, PTA GW backgrounds, and the redshift-dependent merger
rate illustrated in Figure~\ref{fig:merger_rate}—for decisive
confirmation or falsification.

\section{Discussion}
\label{sec:discussion}

\paragraph{QMM information wells vs. curvature–perturbation PBH models.}
Conventional scenarios rely on a transient enhancement of the curvature
power spectrum---for example, through an ultra-slow roll epoch or a
spectator field---to push the radiation‐era density contrast above
$\delta_{\rm c}$ \cite{Ivanov1994,Ashoorioon2019}.  
Those mechanisms require fine-tuning of the inflation potential and risk
generating excessive non-Gaussianity.  
In contrast, \emph{QMM information wells} arise from fundamental
microphysical assumptions: imprint entropy is an unavoidable by-product
of quantum interactions, and its gradients act as pressureless matter.
No dedicated feature in the curvature spectrum is needed; the same
entropy field that solves the black-hole information problem seeds PBHs.

\paragraph{Model parameters.}
Only three knobs control phenomenology:
(i)~the coupling $\lambda$, setting the overall dust fraction;
(ii)~the bounce temperature $T_{\rm B}$, which fixes the $k\!-\!M$
mapping (Eq.\,\ref{eq:k_M_relation}); and
(iii)~the imprint tilt $n_{S}^{(S)}$, which shapes the PBH mass
distribution.  
Current CMB and large-scale structure data already constrain
$\lambda\approx1\pm0.3$ and $n_{S}^{(S)}\lesssim1.4$; PTA and
$\mu$-distortion measurements will soon tighten those bounds.

\paragraph{Open theoretical questions.}
Three issues merit deeper investigation:
\begin{enumerate}
\item \textbf{Entropy decoherence.}  How quickly do imprint states lose
      phase information during a high-curvature bounce, and could that
      alter the post-bounce $P_{S}(k)$?
\item \textbf{Quantum-to-classical transition.}  Does the
      stochastic-classical description of $S(x)$ remain valid once
      gradients become Planckian, or is a fully quantum-gravitational
      treatment required?
\item \textbf{Holographic bounds.}  Because each Planck cell stores a
      finite Hilbert space, is there a maximum imprint energy density,
      and can that cap preempt runaway collapse at extremely blue tilts?
\end{enumerate}
Addressing these questions will clarify whether QMM can serve as a
complete ultraviolet extension of early-universe cosmology.

\section{Conclusions}
\label{sec:conclusions}

We have shown that the Quantum Memory Matrix endows a bouncing universe
with \emph{information wells}—overdensities sourced by gradients in the
imprint-entropy field \(S(x)\).  
These wells grow linearly while outside the horizon, cross the collapse threshold at reentry, and form primordial black holes with a mass spectrum directly traceable to the power-law parameters
\((\lambda,\,n_{S}^{(S)},\,T_{\rm B})\). For \(\lambda\!\sim\!1\) a blue tilt \(n_{S}^{(S)}\!\approx\!1.2\) fills the LIGO–Virgo mass band, whereas slightly steeper spectra yield sub-lunar PBHs that can make up \(\mathcal O(1)\) of dark matter without violating microlensing limits. Because the same imprint stress–energy affects the integrated Sachs–Wolfe signal, \(\mu\)-distortions, and the nanohertz gravitational-wave background, forthcoming CMB-S4, PIXIE, and PTA data will decisively test the scenario. If confirmed, QMM would provide a unified explanation for dark matter, black-hole information retention, and the origin of PBHs—cementing the framework as a cornerstone of the broader QMM program in quantum gravity and cosmology.


\appendix
\setcounter{section}{0}

\renewcommand{\thesection}{\Alph{section}}

\section{Derivation of the QMM Stress–Energy Tensor}
\label{app:stress_energy_derivation}

Starting from
\(
S_{\rm QMM}=\dfrac{\lambda}{2}
            \int d^{4}x\,\sqrt{-g}\,
            g^{\mu\nu}\partial_\mu S\,\partial_\nu S,
\)
vary the action with respect to the metric:
\[
\delta S_{\rm QMM}
 =\frac{\lambda}{2}\int\! d^{4}x
  \Bigl[
    \sqrt{-g}\,\partial_\mu S\,\partial_\nu S\,\delta g^{\mu\nu}
   -\tfrac12\sqrt{-g}\,g^{\mu\nu}\partial_\mu S\,\partial_\nu S\,g_{\alpha\beta}\,\delta g^{\alpha\beta}
  \Bigr].
\]
Using \(\delta(\sqrt{-g})=-\tfrac12\sqrt{-g}\,g_{\alpha\beta}\,\delta g^{\alpha\beta}\)
and lowering indices,
\[
\delta S_{\rm QMM}
 =-\frac12\!\int\! d^{4}x\,\sqrt{-g}\;
   T_{\mu\nu}^{(\mathrm{QMM})}\,\delta g^{\mu\nu},
\qquad
T_{\mu\nu}^{(\mathrm{QMM})}
 =\lambda
   \bigl[
     \partial_\mu S\,\partial_\nu S
    -\tfrac12 g_{\mu\nu}(\partial S)^2
   \bigr].
\]
Because \(S\) is a \emph{dynamical} scalar, the Belinfante procedure
allows one to add the identically conserved improvement term
\(\lambda(g_{\mu\nu}\Box S-\nabla_\mu\nabla_\nu S)\), yielding
Eq.\,( \ref{eq:TmunuQMM} ) in the main text.  Conservation follows
directly from the Klein–Gordon equation \(\Box S=0\).

\section{Constraints on the Imprint Power Spectrum}
\label{app:power_spectrum}

\begin{figure}[t]
  \centering
  \includegraphics[width=0.75\textwidth]{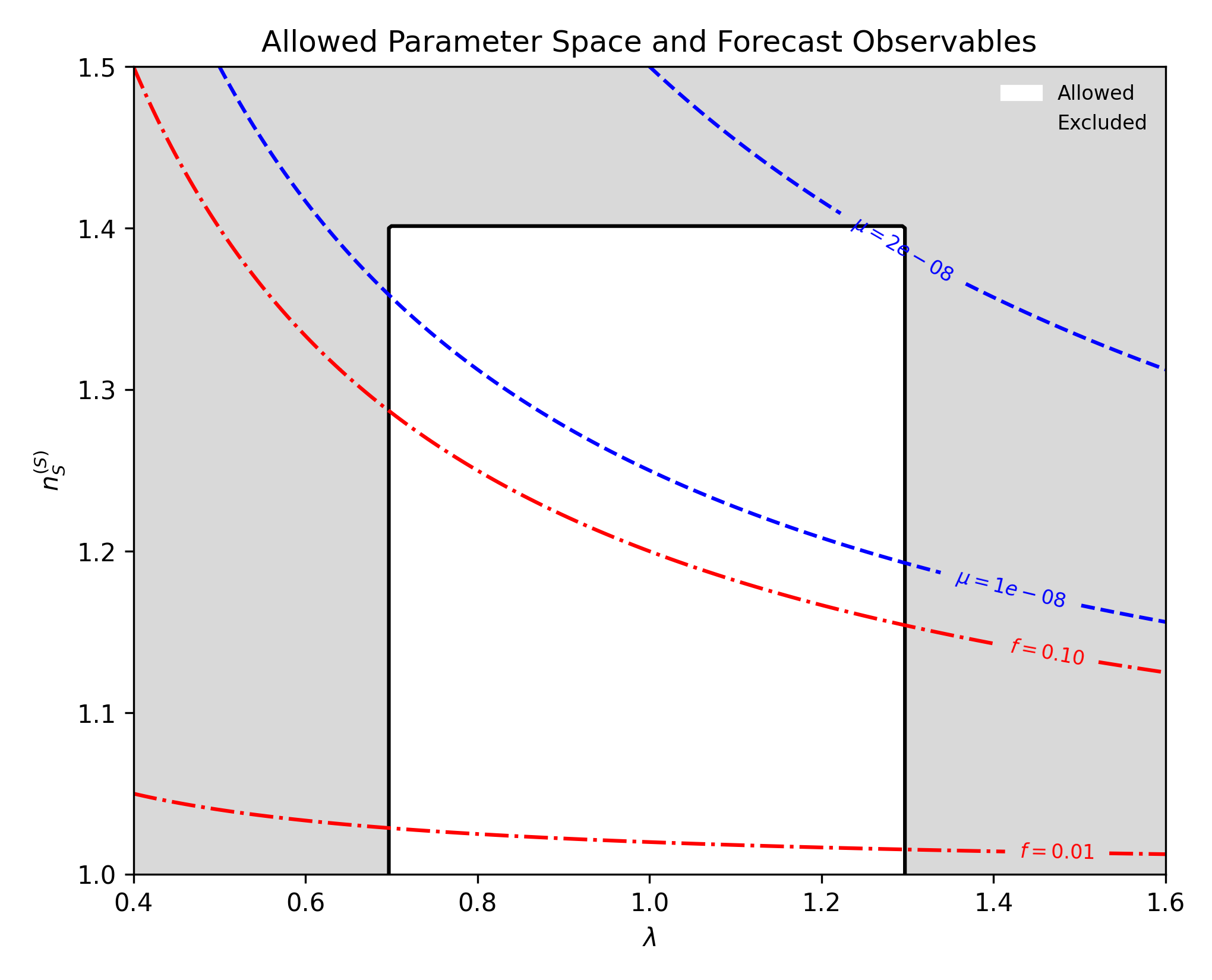}
  \caption{Allowed (white) and excluded (gray) regions in the
           $(\lambda,n_{S}^{(S)})$ plane after \textit{Planck}+BAO+LSS
           constraints.  Blue dashed contours indicate constant
           $\mu$-distortion; red dash–dot contours show the present-day
           PBH fraction $f_{\rm PBH}$.}
  \label{fig:param_space}
\end{figure}

\textit{Planck} TT+TE+EE+lowE+lensing data, combined with BOSS BAO,
limit any dust‐like component prior to matter–radiation equality to
\(\Omega_{\rm QMM}h^{2}\le0.003\) (95\,\% CL) \cite{Planck2018}.  
Because \(\Omega_{\rm QMM}\propto\lambda A_{S}\) this implies
\[
0.7\;\lesssim\;\lambda\,A_{S}^{1/2}\;\lesssim\;1.3
\quad(1\sigma).
\]
Large-scale-structure power spectra from eBOSS and DES further restrict
the blue tilt to \(n_{S}^{(S)}\lesssim1.4\) for
\(k\le1\,h\,\mathrm{Mpc}^{-1}\).  
These constraints are summarized in Figure~\ref{fig:param_space}, which
plots the allowed region in the \((\lambda,n_{S}^{(S)})\) plane and
overlays contours of constant $\mu$-distortion and present-day PBH
fraction.  The figure highlights in particular the sub‐regions that
yield (i) $\gtrsim10\%$ PBH dark matter, (ii) $\gtrsim1\%$ ISW excess,
and (iii) $\mu>3\times10^{-8}$.

\section{PBH Mass Function Details}
\label{app:mass_function}

The variance of the density contrast smoothed on scale \(R\) is
\[
\sigma^{2}(R)=\lambda^{2}\!\int\! \frac{dk}{k}\,
              P_{S}(k)\,W^{2}(kR),
\]
where
\(W(kR)=3[\sin(kR)-kR\cos(kR)]/(kR)^{3}\) is the real‐space top‐hat
window.  
Critical collapse modifies the monochromatic approximation:
\[
M=\kappa\,M_{\rm H}
  \bigl(\delta-\delta_{\rm c}\bigr)^{\gamma_{\rm crit}},
\qquad
\kappa\simeq3.3,\;
\gamma_{\rm crit}\simeq0.36.
\]
Transforming variables,
\[
\beta(M)=\int_{\delta_{\rm c}}^{\infty}\!
         \frac{d\delta}{\sqrt{2\pi}\,\sigma}\,
         \exp\!\Bigl[-\tfrac{\delta^{2}}{2\sigma^{2}}\Bigr]\,
         \delta\!\bigl(M-\kappa M_{\rm H}
                          (\delta-\delta_{\rm c})^{\gamma_{\rm crit}}\bigr).
\]
A saddle-point expansion around \(\delta_{\rm c}\) reproduces
Eq.\,( \ref{eq:beta_approx} ) in leading order.  
The notebook computes \(\beta(M)\), \(\Omega_{\rm PBH}(M)\), and
plots $f_{\rm PBH}$ against observational limits.

\section{Numerical Setup for Bounce–to–PBH Simulations}
\label{app:numerics}

\subsection{\texorpdfstring{Modified Boltzmann Code (\textsc{CLASS}/\textsc{CAMB})}{Modified Boltzmann Code (CLASS/CAMB)}}

\begin{enumerate}
\item \textbf{New fluid.}  
      Add a pressureless species \texttt{qmm} with energy density
      \(\rho_{\rm QMM}(a)=\lambda(\dot S^{2}+(\nabla S)^{2}/a^{2})/2\).
\item \textbf{Perturbations.}  
      Integrate Eqs.\,( \ref{eq:delta_QMM_eom} ) for each $k$ in
      \texttt{perturb\_sources.c}, sourcing metric potentials alongside
      CDM and baryons.
\item \textbf{Initial conditions.}  
      Sample $P_{S}(k)$ at the bounce scale factor $a_{\rm B}$, rescale
      to $a_{\rm init}=10^{-8}$, and impose
      \(\delta_{\rm QMM}(k,a_{\rm init})=C_{1}(k)\,a_{\rm init}/a_{\rm B}\).
\item \textbf{Background module.}  
      Insert $\rho_{\rm QMM}(a)$ into the Friedmann integrator; ensure
      the stiff term $\dot S^{2}$ is updated with a Runge–Kutta wrapper.
\item \textbf{Validation.}  
      Recover $\Lambda$CDM spectra when \(\lambda\to0\) and replicate
      CAMB’s $\ell$–spectra within $0.1\%$ for standard parameters.
\end{enumerate}

\subsection{N-Body with Live \texorpdfstring{$S$}{S}-Grid}

\begin{itemize}
\item \textbf{Code base.}  
      Fork \textsc{Gadget-4} to include a lattice field $S$ stored on the
      same PM grid as the density assignment.
\item \textbf{Equations of motion.}  
      Advance particles with
      \(\ddot{\bm x}=-\nabla(\Phi+\Phi_{\rm QMM})\),
      where $\nabla^{2}\Phi_{\rm QMM}=4\pi G\rho_{\rm QMM}$; update
      \(\rho_{\rm QMM}\) each step via
      \(\rho_{\rm QMM}=\lambda[(\nabla S)^{2}+a^{2}\dot S^{2}]/2\).
\item \textbf{Grid resolution.}  
      $N_{\rm grid}=2048^{3}$ for a $L=500\,h^{-1}{\rm Mpc}$ box gives
      $\sim100$ cells per comoving PBH Jeans length at
      $T_{\rm re}=10^{5}\,{\rm GeV}$.
\item \textbf{Time stepping.}  
      Use a global $\Delta a=2\times10^{-3}$ until
      $a=10^{-2}$, then individual particle and field time steps
      satisfying $\Delta t\le0.02H^{-1}$.
\item \textbf{Convergence tests.}  
      Halve $\Delta a$, double $N_{\rm grid}$, and confirm PBH mass
      function changes $\le5\%$.  Compare with one-dimensional
      high-resolution collapse to verify critical exponent recovery.
\end{itemize}

\bigskip
\noindent\textbf{Supplementary notebook.} 
All numerical routines, Boltzmann‐code patches, and plotting commands
are provided in the Jupyter file available with the source files of
this article. Executing the notebook regenerates Figures \ref{fig:PS_tilt} - \ref{fig:param_space},
reproduces the $(\lambda,n_{S}^{(S)})$ constraints in
Appendix~\ref{app:power_spectrum}, and outputs machine-readable tables of the PBH mass function and merger-rate predictions.




\end{document}